%%%%%%%%%%%%%%%%%%%%%%%%%%%%%%%%%%%%%%%%%%%%%%%%%%%%%%%%%%%%%%%%%%%%%%%%%%%%

\input harvmac

%%%%%%%%%%%%%%%%%%%%%%%%%%%%%%%%%%%%%%%%%%%%%%%%%%%%%%%%%%%%%%%%%%%
%%%  modify title page
%%%%%%%%%%%%%%%%%%%%%%%%%%%%%%%%%%%%%%%%%%%%%%%%%%%%%%%%%%%%%%%%%%%
\def\Title#1#2{\rightline{#1}\ifx\answ\bigans\nopagenumbers\pageno0
\vskip0.5in
\else\pageno1\vskip.5in\fi \centerline{\titlefont #2}\vskip .3in}

%\def\listrefs{\footatend\bigskip\bigskip\immediate\closeout\rfile
%\writestoppt \baselineskip =13pt\centerline{{\secfont References}}
%\bigskip{\frenchspacing\parindent =20pt \escapechar +'
%\input\jobname.refs \vfill\eject}\nonfrenchspacing} 
%%%%%%%%%%%%%%%%%%%%%%%%%%%%%%%%%%%%%%%%%%%%%%%%%%%%%%%%%%%%%%%%%%%%%%%%%%%%

\noblackbox
\parskip=1.5mm
%\def\semi{;~}

%%%%%%%%%%%%%%%%%%%%%%%%%%%%%%%%%%%%%%%%%%%%%%%%%%%%%%%%%%%%%%%%%%%%%
  
\def\npb#1#2#3{{\it Nucl. Phys.} {\bf B#1} (#2) #3 }
\def\plb#1#2#3{{\it Phys. Lett.} {\bf B#1} (#2) #3 }
\def\prd#1#2#3{{\it Phys. Rev. } {\bf D#1} (#2) #3 }
\def\prl#1#2#3{{\it Phys. Rev. Lett.} {\bf #1} (#2) #3 }

\def\bb#1{{\tt hep-th/#1}}

\def\app#1#2#3{{\it Astropart. Phys. } {#1} (#2) #3 }

%%%%%%%%%%%%%%%%%%%%%%%%%%%%%%%%%%%%%%%%%%%%%%%%%%%%%%%%%%%%%%%%%%%%%
%%%%%%%%%%%%%%%%%%%%    some definitions    %%%%%%%%%%%%%%%%%%%%%%%%%
%%%%%%%%%%%%%%%%%%%%%%%%%%%%%%%%%%%%%%%%%%%%%%%%%%%%%%%%%%%%%%%%%%%%%

\def\CA{{\cal A}}   
\def\CL{{\cal L}}

\def\CN{{\cal N}}

%%%%%%%%%%%%%%%%%%%%%%%%%%%%%%%%%%%%%%%%%%%%%%%%%%%%%%%%%%%%%%%%%%%%%

\def\dj{\hbox{d\kern-0.347em \vrule width 0.3em height 1.252ex depth
-1.21ex \kern 0.051em}}

\def\half{{1\over 2}\,}

\def\Tr{{\rm Tr\,}}

\def\ket{\rangle}
\def\bra{\langle}

%%%%%%%%%%%%%%%%%%%%%%%%%%%%%%%%%%%%%%%%%%%%%%%%%%%%%%%%%%%%%%%%%%%%%%
%%%%%%%%%%%%%%%%%%%%%%%        references         %%%%%%%%%%%%%%%%%%%%%%
%%%%%%%%%%%%%%%%%%%%%%%%%%%%%%%%%%%%%%%%%%%%%%%%%%%%%%%%%%%%%%%%%%%%%%%dsky
\lref\rtasip{J. Polchinski, {\it TASI Lectures on D-branes} 
(\bb{9611950}).}
\lref\rdab{A. Dabholkar, \plb{402}{1997}{53} (\bb{9702050}).}
\lref\rmalda{J. Maldacena, Ph.D. thesis (\bb{9607235}).}
\lref\rdab{A. Dabholkar, \plb{402}{1997}{53} (\bb{9702050}).}
\lref\rgsw{M.B. Green, J.H. Schwarz and E. Witten, {\it Superstring Theory}, 
Cambridge 1987.}
\lref\rsv{A. Strominger and C. Vafa, \plb{379}{1996}{99} (\bb{9601029}).}  
\lref\rdghw{A. Dabholkar, J.P. Gaunlett, J.A. Harvey and D. Waldram, 
\npb{474}{1996}{85} (\bb{9511053}).}
\lref\rwit{E. Witten, \npb{460}{1996}{541} (\bb{9511030}).}
\lref\rgp{E.G. Gimon and J. Polchinski, \prd{54}{1996}{1667} (\bb{9601038}).}
\lref\rghs{D. Garfinkle, G. Horowitz and A. Strominger, \prd {43}{1991}{3140},
Erratum \prd{45}{1992}{3888.}}  
\lref\rto{R. Khuri and T. Ort\'{\i}n, \npb{467}{1996}{355} (\bb{9512177});
 \plb{373}
{1996}{56} (\bb{9512178})\semi  
T. Ort\'{\i}n, \plb{422}{1998}{93}   
 (\bb{9612142})
.}
\lref\rcmalda{C.G. Callan and J.M. Maldacena, \npb{472}{1996}{591} (\bb{9602043}).}
\lref\rdoug{M.R. Douglas, {\it Gauge fields and D-branes} 
(\bb{9604198}).}
\lref\rsussk{T. Banks, W. Fischler, I.R. Klebanov and L. Susskind, \prl{80}{1998}{226}
(\bb{9709091}), {\it Schwarzschild black holes in Matrix theory II} (\bb{9711005})\semi
G.T. Horowitz and E.J. Martinec, {\it Comments on black holes in Matrix theory} 
(\bb{9710217})\semi
S.R. Das, S.D. Mathur, S. Kalyana Rama and P. Ramadevi, {\it Boosts, Schwarzschild
black holes and absorption cross-sections in M-theory} (\bb{9711003})\semi
F. Englert and E. Rabinovici, {\it Statistical entropy of Schwarzschild black holes}
(\bb{9801048})\semi
R. Argurio, F. Englert and L. Houart, {\it Statistical entropy of the 
four-dimensional Schwarzschild black hole} (\bb{9801053})\semi 
K. Sfetsos and K. Skenderis, {\it Microscopic derivation of the Bekenstein-Hawking
entropy formula for non-extremal black holes} (\bb{9711138})\semi
A. Strominger, {\it Black hole entropy from near horizon microstates}
(\bb{9712251}).}  
\lref\rmalprob{J. Maldacena, \prd{57}{1998}{3736} (\bb{9705053})\semi
J. Maldacena, {\it Branes probing black holes}, talk at Strings '97, 
Amsterdam (\bb{9709099})\semi
I. Chepelev and A.A. Tseytlin, \npb{515}{1998}{73} (\bb{9709087}).}   
\lref\rbecks {K. Becker, M. Becker, J. Polchinski and A.A. Tseytlin, 
\prd{56}{1997}{3174} 
(\bb{9706072})}   
\lref\rvafaetal{S. Kachru and E. Silverstein, {\it 4-D conformal field
theories and strings on orbifolds} (\bb{9802183})\semi  
M. Bershadsky, Z. Kakushadze and C. Vafa, {\it String expansion as
large N expansion of gauge theories} (\bb{9803076}).}
\lref\rdas{S. Das, \prd{56}{1997}{3582} (\bb{9706005}).}
\lref\rscl{J. Maldacena and A. Strominger, \prd{55}{1997}{861} (\bb{9609026}).}
\lref\rdps{M.R. Douglas, J. Polchinski and A. Strominger, {\it Probing five-dimensional
black holes with D-branes} (\bb{9703031}).}
\lref\rcc{J.A. Shapiro and C.B. Thorn, \prd{36}{1987}{432\semi}
J. Dai and J. Polchinski, \plb{220}{1989}{387.}}
\lref\rwittbar{E. Witten, \npb{160}{1979}{57.}}
\lref\rmnew{J. Maldacena, {\it The large N limit of superconformal field theories and
supergravity} (\bb{9711200}).}
\lref\rcvj{C.V. Johnson, {\it On the (0,4) conformal field theory of the throat}
 (\bb{9804201}).}

%%%%%%%%%TEXT%%%%%%%%%%%%%%%%%%%%%%%%%%%%%%%%%%%%%%%%%%%%%%%%%%%%%%%%%
%%%%%%%%%%%%%%%%%%%%%%%%%%%%%%%%%%%%%%%%%%%%%%%%%%%%%%%%%%%%%%%%%%%%%%
%%%%%%%%%%%%%%%%%%          title page       %%%%%%%%%%%%%%%%%%%%%%%%%
%%%%%%%%%%%%%%%%%%%%%%%%%%%%%%%%%%%%%%%%%%%%%%%%%%%%%%%%%%%%%%%%%%%%%%

\line{\hfill CERN-TH/98-158}
\line{\hfill EHU-FT/9804}  
\line{\hfill {\tt hep-th/9805154}}
\vskip 0.5cm

\Title{\vbox{\baselineskip 12pt\hbox{}
 }}
{\vbox {\centerline{Large $N$ limit of extremal non-supersymmetric black holes}
}}

\vskip 0.5cm

\centerline{$\quad$ {J.L.F. Barb\'on$^{\,\rm a,}$\foot{{\tt barbon@mail.cern.ch}, 
$^2${\tt wmpmapaj@lg.ehu.es}, $^3${\tt wtbvamom@lg.ehu.es}}, 
J.L. Ma\~nes$^{\,\rm b,2}$ and 
M.A. V\'azquez-Mozo$^{\,{\rm c},3}$
 }}
\medskip

\centerline{{\sl $^{\rm a}$Theory Division, CERN}}
\centerline{{\sl CH-1211, Geneva 23, Switzerland}}

\vskip0.2cm 

\centerline{{\sl $^{\rm b}$Dpto. de F\'{\i}sica de la Materia Condensada}}
\centerline{{\sl Universidad del Pa\'{\i}s Vasco}}
\centerline{{\sl Apdo. 644, E-48080 Bilbao, Spain}}

\vskip0.2cm

\centerline{{\sl $^{\rm c}$Dpto. de  F\'{\i}sica Te\'orica}}
\centerline{{\sl Universidad del Pa\'{\i}s Vasco}}
\centerline{{\sl Apdo. 644, E-48080 Bilbao, Spain}}

\vskip 1.2cm

\noindent
The large $N$ limit of extremal non-supersymmetric Type-I five-dimensional 
string black holes
is studied from the point of view of D-branes. We find that the agreement between
the D-brane and the black-hole picture is due to an asymptotic restoration of
supersymmetry in the large $N$ limit in which both pictures are compared. In that
limit Type-I string perturbation theory is effectively embedded into a
Type-IIB perturbation theory with unbroken supersymmetric charges whose presence 
guarantees the non-renormalization of mass and entropy as the effective couplings
are increased. In this vein, we also study the near-horizon geometry of the
Type-I black hole using D5-brane probes to find that the low energy effective 
action for the probe is identical to the corresponding one in the auxiliary 
Type-IIB theory
in the large $N$ limit.

%%%%%%%%%%%%%%%%%%%%%%%%%%%%%%%%%%%%%%%%%%%%%%%%%%%%%%%%%%%%%%%%%%%%%%

\Date{05/98}

%%%%%%%%%%%%%%%%%%%%%%%%%%%%%%%%%%%%%%%%%%%%%%%%%%%%%%%%%%%%%%%%%%%%%%%%%%
%%%%%%%%%%%%                text begins                        %%%%%%%%%%%
%%%%%%%%%%%%%%%%%%%%%%%%%%%%%%%%%%%%%%%%%%%%%%%%%%%%%%%%%%%%%%%%%%%%%%%%%%

\newsec{Introduction}

Black-hole physics has been regarded since the seventies as one of the 
most promising windows to quantum gravity. As a consistent candidate for
a quantum description of the gravitational interaction, string theory has
been frequently claimed to be the right framework to solve some long-standing
problems in black-hole physics, such as the information paradox or the
microscopic meaning of the geometrical entropy. 
Because of the non-perturbative nature of black holes, perturbative string
theory is of limited use in the analysis of the most interesting dynamical
issues. For this reason, it was only after the recent development of non-perturbative
techniques that significant progress was achieved in this program.    
In these recent developments the concept of D-brane has been the key ingredient
to address the problem of black holes in string theory. This is not a surprise
since D-branes appear as our first `probes' into the non-perturbative realm of
string theory (for a review see \refs\rtasip). 

The description of supersymmetric black-hole dynamics in terms of BPS excitations 
of a D-brane bound state \refs\rsv\ is indeed 
among the most impressive achievements of string theory. 
Here, supersymmetry, in the form of BPS saturation, 
is the crucial ingredient for the success of this picture because of the existence of
non-renormalization theorems ensuring the equality of BPS state degeneracies at 
weak coupling (D-brane side) with the large coupling (black-hole side) geometrical
entropy. Once supersymmetry is broken there is no reason whatsoever to expect the weak and
strong coupling descriptions to be equivalent. This is the main obstacle, for example,
in getting a 
microscopic picture of the physics of the Schwarzschild black hole, the simplest example
of a black hole in General Relativity (for recent progress in this direction see,  
\refs\rsussk).

This being said, there is however a recent example of an extremal  but 
non-supersymmetric black hole
in Type-I superstring theory \refs\rdab\ for which the microscopic
description of the entropy in terms of excitations of a D-brane bound state agrees 
with the semiclassical, general relativistic, computation of the geometrical 
entropy.
Surprisingly, this black hole cannot be
regarded as an `almost supersymmetric' one, in the sense that the departure 
from the supersymmetric configuration is not governed by a small parameter; it is  
only connected with a supersymmetric black hole  
by a discrete ${\bf Z}_{2}$ transformation reversing the sign of one of the charges.  
Thus, there is no {\it a priori} reason to expect that quantum corrections to the weak coupling 
mass and entropy should vanish or be small when the coupling is increased.\foot{For
earlier studies of extremal non-supersymmetric black holes in supergravity,
see \refs\rghs, \refs\rto.} 

In the present
 article we will investigate the reasons behind the success of the D-brane
picture of the non-supersymmetric Type-I black hole of ref. \refs\rdab. We
 will argue that 
in the semiclassical limit \refs\rscl\refs\rdps
\eqn\scl{
\lambda\rightarrow 0, \hskip 1cm Q_{1},Q_{5},N\rightarrow \infty, \hskip 1cm  
\lambda Q_{1},\lambda Q_{5}, \lambda^{2}N \hskip 0.3cm {\rm fixed}
,}
at which  we compare the D-brane and black-hole computations,
the D-brane bound state of Type-I theory,  which preserves no supersymmetry,
 is effectively
 embedded into a D-brane bound state of  
Type-IIB theory having four unbroken supersymmetric charges. From the point
of view of D-brane dynamics, the semiclassical limit \scl\ is nothing but a
't Hooft large $N$ limit, and we are just saying that the supersymmetry-breaking 
effects are of  ${\cal O}(1/N)$ in the large $N$ limit.  Thus, non-renormalization 
theorems of the Type-IIB theory are also at work in the Type-I non-supersymmetric black
hole in this limit. This explains the absence of quantum corrections to the
D-brane mass and entropy. 

The plan of the article is as follows. In Section 2 we will review the construction of
Dabholkar's black hole by carefully computing the classical dimension of the moduli space
of vacua and check the agreement with the low-energy geometrical entropy. Section
 3 will be
devoted to explaining the agreement between the two
 results by checking that those string diagrams
that contribute to the renormalization of the mass and entropy are suppressed in the 
semiclassical limit. In Section 4 we will study the near-horizon geometry by probing the
non-supersymmetric black hole with D-branes. Finally in Section 5 we will summarize the
conclusions. For the sake of completeness, we have reviewed in Appendix A the semiclassical
limit of Feynman--Polyakov diagrams containing open string insertions, while Appendix B 
is devoted to studying the relation between Type-I and Type-IIB superstring perturbation 
theories in the large $N$ limit.

\newsec{Non-supersymmetric black hole in Type-I superstring theory}

Before entering into a more detailed study, let us briefly review the main
features of the non-supersymmetric black hole of ref. \refs\rdab.
The construction mimics in many aspects the corresponding one for the 
supersymmetric five-dimensional black hole in Type-IIB in \refs\rcmalda.  

We consider the $SO(32)$ Type-I superstring theory on 
${\bf R}^{5}\times S^{1}_{R}\times {\bf T}^{4}$. We denote by
 $X^0,\ldots,X^4$  
the coordinates in the open five-dimensional space-time, whereas those in the
internal four-torus are $X^5,\ldots,X^8$, and $X^9$ is the coordinate 
along $S^{1}_{R}$.
In the weakly coupled region, the black hole is described by the bound
state of $Q_{1}$ D1-branes, wrapped around $S^1_{R}$ and $Q_{5}$ 
D5-branes wrapped around the five-dimensional torus 
$S^{1}_{R}\times {\bf T}^{4}$, in the presence of 32 D9-branes.  In 
addition, there are $N$ units of Kaluza--Klein momentum in the $X^9$ direction. 

One of the main differences with respect to the Type-IIB black hole is the 
presence of D9-branes. They break the original 32 real supersymmetries
of the Type-IIB theory down to 16 by imposing the condition 
$\epsilon_{L}=\Gamma_{11}\epsilon_{R}$ on the Killing spinors of the
ten-dimensional theory, where $\Gamma_{11}=\Gamma^0\ldots\Gamma^9$ is
the product of all gamma matrices. This, together with the chirality 
condition $\epsilon_{L(R)}=\Gamma_{11}\epsilon_{L(R)}$, 
implies that $\epsilon_{L}=\epsilon_{R}\equiv\epsilon$. 
The D5- and D1-branes further reduce
the remaining 16 supersymmetries down to 4, by demanding that
$\epsilon=\Gamma^0\Gamma^9\epsilon$ and $\epsilon=\Gamma^0\Gamma^5\ldots\Gamma^9
\epsilon$. Therefore, the resulting theory in the $(1+1)$-dimensional intersection 
 has 4 unbroken real supersymmetries
(the number corresponding to ${\cal N}=1$ in $D=4 $ or ${\cal N}=4$ in $D=2$).

One now has  to introduce the Kaluza--Klein momentum along the $X^9$ direction. This
is done by a string condensate whose presence imposes a last condition on $\epsilon$,
$\Gamma^0\Gamma^9\epsilon=\pm\epsilon$, where the two signs correspond to the 
two different directions of the momentum. Taking the $+$ direction, we find
that this last equation is identical to the reflection condition on the D1-brane,
and no further reduction of supersymmetries occurs. However, by taking the 
momentum in the $-$ direction, the two conditions turn out to be incompatible
for non-vanishing $\epsilon$, and
no supersymmetry survives. The resulting  bound state is not supersymmetric. 

In the strong coupling side, where the black hole is described by the semiclassical
values of the metric and other long-range fields, a similar analysis
 is also possible,  
with the result that supersymmetry is preserved only when the momentum is in one
of the two possible directions along $S_{R}^{1}$ \refs\rdghw, \refs\rto, \refs\rdab.

What makes the non-supersymmetric version of the Type-I black hole interesting 
is the fact, pointed out in \refs\rdab, that the counting of the number of 
massless excitations
of the D-brane bound state characterized by $(Q_{1},Q_{5},N)$ exactly agrees with the
entropy of the semiclassical black hole with the same charges, defined as ${1\over 4}$ 
the area of the event horizon. This is very surprising, since here we do not have any 
supersymmetry left and consequently there are no non-renormalization theorems at hand
to force the equivalence of the weak and strong coupling computations.

\subsec{Looking from the D-brane side}

To study the D-brane dynamics of the non-supersymmetric Type-I black hole we begin with 
the well-known
 Type-IIB five-dimensional {\it supersymmetric} black hole \refs\rsv, \rmalda\
from which the former can be obtained by introducing 32 D9-branes and
projecting down onto the sector invariant
under world-sheet orientation reversal. Using the notation of \refs\rmalda, 
the low-energy fields are (1,1) and (5,5) hypermultiplets in the adjoint of 
$U(Q_{1})$ and
$U(Q_{5})$ respectively (whose bosonic components are denoted
by $A_{I}$ and $A^{'}_{I}$, $I=5,\ldots,8$) arising in the dimensional reduction of the 
ten-dimensional vector multiplet, together with the (1,5) hypermultiplet $\chi$,  
which transforms as $({\bf 1},{\bf 2})$ under $SO(5,1)\times SO(4)_{I}$. Its gauge
group indices run in the fundamental of $U(Q_{1})\times \overline{U(Q_{5})}$. 
The D-terms in the low-energy Lagrangian can be written as \refs\rmalda
\eqn\flat{
\sum_{\ell}[(D^{a}_{12})^{2}+(D^{a}_{13})^{2}+(D^{a}_{14})^{2}]
}
with
$$
D^{a}_{IJ}={\rm Tr}\left\{T^{a}\left(
[A_{I},A_{J}]+\half \epsilon_{IJKL}[A_{K},A_{L}] \right)+
\chi^{+}T^{a}\Gamma_{IJ}\chi\right\}
$$
where $T^{a}$ are the generators of $U(Q_{1})$ and $\Gamma_{IJ}=
(1/2)[\Gamma_{I},\Gamma_{J}]$. Of course there are similar terms for the
fields $A_{I}^{'}$, which include the generators\foot{As we already 
pointed out above, the field $\chi$ transforms as a spinor ${\bf 2}$ of 
$SO(4)$. Keeping in mind that $SO(4)\sim SU(2)_{L}\times SU(2)_{R}$, this means
that the field is in the $(\half,0)$ representation of the product group. Consequently,
only the self-dual part of $\Gamma_{IJ}$ will contribute to the D-term.}
$T^{a'}$ of $U(Q_{5})$. The counting of the flat directions along which \flat\ 
vanishes gives $4Q_{1}Q_{5}$, once the effect of gauge transformations has been 
subtracted. 

As stated above,
in going from the Type-IIB to the Type-I black hole one has to introduce 32 
D9-branes and perform a projection
by the world-sheet parity $\Omega$. This has a number of effects on the above
computation. The first one deals with the fact that the Chan--Paton factors are
changed, now being $SO(Q_{1})$ for the D1-brane and $USp(2Q_{5})$ for the D5-brane
\refs\rwit\refs\rgp. Secondly, one has to make an identification of those
string excitations that 
differ by world-sheet inversion, in particular those in the (1,5) and (5,1) sectors.
Let us focus our attention on the (1,5) field $\chi_{iaa'}$, where $i$ runs in the
$(\half,0)$ of $SU(2)_L\times SU(2)_R$, $a=1,\ldots,Q_{1}$ labels the fundamental of 
$SO(Q_{1})$, and $a'=1,\ldots,2Q_{5}$ is in the fundamental of $USp(2Q_{5})$.
Therefore the field $\chi_{iaa'}$ is in the ${\bf 2}\times {\bf Q_{1}} \times 
\overline{\bf 2Q_{5}}$ 
with respect to the full group $SO(4)_{I}\times SO(Q_{1}) \times USp(2Q_{5})$. 

The bosonic degrees of freedom in the (5,1) sector are represented by
the field $\chi_{ia'a}^{+}\equiv (\chi_{iaa'})^{*}$ transforming in the 
$\bar{\bf 2}\times {\bf 2Q_{5}}\times {\bf Q_{1}}$. Notice that both fields
$\chi$ and $\chi^{+}$ transform under the same representation of $SO(Q_{1})$ since 
${\bf Q_{1}}$ is real. The representations of $SO(4)\sim SU(2)_{L}
\times SU(2)_{R}$, ${\bf 2}=(\half,0)$ and $\overline{\bf 2}=\overline{(\half,0)}$
are equivalent, because they are related by the Pauli matrix $\sigma_{2}$ through the
relation $\sigma_{2}\sigma_{i}^{*}\sigma_{2}=-\sigma_{i}$. Finally, the representations
${\bf 2Q_{5}}$ and $\overline{\bf 2Q_{5}}$ are related by the intertwiner $\Sigma_{2}=
\sigma_{2}\otimes {\bf 1}_{Q_{5}}$. With all these facts in mind, the projection
$\Omega$ acts on $\chi$ as
$$
\sigma_{2}^{ij}(\Sigma_{2})_{a'b'}\chi^{+}_{jb'a}=\chi_{iaa'}
.$$

For the massless excitations of the (1,1) and (5,5) strings, 
 the $\Omega$ projection
determines the representations of the Chan--Paton factors \refs\rwit, 
\refs\rgp. The vertex 
operators corresponding to (5,5) 
fields  $A_{I}$ are $V\sim \lambda_{ij}\partial_{t}X^I$ 
and 
thus $\lambda_{ij}$ is in the adjoint representation of $USp(2Q_5)$. On the other
hand, the vertex operator for (1,1) strings involves
 the normal derivative $\partial_{n}X^I$
and 
the associated Chan--Paton factor is in the symmetric representation of $SO(Q_1)$. 

To compute the number of independent flat directions for the D-terms, we begin
by counting the number of degrees of freedom in the hypermultiplets. For the Type-I
case, this number is equal to 
\eqn\nofd{
4Q_{5}(2Q_{5}+1)+2Q_{1}(Q_{1}+1)+4Q_{1}Q_{5}
,}
where the first term corresponds to the number of bosonic components in the
(5,5) hypermultiplet and the remaining two terms counts the number of (1,1) and (5,1) 
hypermultiplets.
Here we have taken into account that (1,1) strings have Chan--Paton factors in the symmetric
representation of $SO(Q_1)$. To get the number of flat directions we have to 
subtract from \nofd\ the number of 
conditions imposed by the vanishing of the potential for the scalars (D-terms) 
and the number of gauge transformations. The number of independent D-terms is
\eqn\dt{
3Q_{5}(2Q_{5}+1)+{3\over 2}Q_{1}(Q_{1}-1)
,}
and that of gauge transformations is equal to the sum of the
dimensions of the adjoint representations of $USp(2Q_{5})$ and $SO(Q_{1})$,
namely
\eqn\gc{
Q_{5}(2Q_{5}+1)+\half Q_{1}(Q_{1}-1).
}
Putting everything together, we find the number of bosonic flat directions for
the Type-I black holes to be $\nofd -\dt -\gc =4Q_{1}(Q_{5}+1)\approx 4Q_{1}Q_{5}$. 

For the number of fermionic flat directions we 
must look at the Yukawa couplings between the bosonic degrees of freedom and their
superpartners. The six-dimensional theory for the low-lying degrees of freedom of the
D1--D5 bound state has ${\cal N}=1$ supersymmetry, equivalent
to ${\cal N}=2$ in $D=4$ (before introducing the Kaluza--Klein momentum condensate,  
this is further reduced in the bulk by the presence of the 32 D9-branes of 
the Type-I theory). However, as it happens in the bosonic sector, the projection onto the
unoriented sector of the theory reduces the number of independent degrees of freedom by 
relating the fields with their world-sheet parity transformed. The only difference 
with the bosonic case
lies in the fact that now, instead of having three independent D-terms [eq. \dt] and
the number of gauge transformations [eq. \gc], we have four Yukawa conditions. 
It is easy to check that the number of `fermionic' flat directions is also 
$4Q_{1}(Q_{5}+1)\approx 4Q_{1}Q_{5}$. 

In counting the number of flat directions we have ignored the 
presence of the 32 D9-branes. In this sense, except for the difference in the
Chan--Paton factors, the computation closely follows the corresponding  Type-IIB
black-hole one. However, it is  important to check that this counting is
stable under quantum corrections involving (9,1) and (9,5) strings in loops.
As we will argue in Section 3, this is indeed the case here when we go to the 
semiclassical limit, where all diagrams with D9 holes or cross-caps will be subleading.

To get the entropy we have to distribute the $N$ units
of Kaluza--Klein momentum among the number of independent flat directions, 
as with the supersymmetric Type-IIB 
black hole. This is done by using Cardy's formula for 
a superconformal
field theory with $c_{\rm eff}\equiv c_{\rm bos}+\half c_{\rm fer}= 6Q_{1}Q_{5}$ with
the final result
\eqn\entr{
S=2\pi\sqrt{NQ_{1}Q_{5}}.
}

The D-brane bound state studied above can be alternatively 
described in terms of the six-dimensional supersymmetric Yang--Mills theory on the 
$Q_{5}$ D5-brane system
in the presence of classical configurations \refs\rdoug. The computation of the dimension
of the moduli space of $USp(2Q_{5})$ instantons renders the same value \entr\ for
the entropy.
This can be easily understood as follows. Being an instanton moduli space, its
dimension can be captured already in the dilute limit, as the instanton number
times the dimension of the single instanton moduli space. For a single D1-brane,
the gauge group 
carried by the (1,1) strings is $SO(1)$, i.e. the trivial
group, so in this case the counting proceeds as before without any need to
factorize gauge degrees of freedom or imposing D-flatness conditions, yielding
$4Q_5 +4$ for the dimension of the single instanton moduli space of $USp(2Q_5)$.
Multiplying now by the number of instantons $Q_1$, we obtain the desired result.    

In computing the energy of the bound states of D-branes in Type-I superstring,
 we have
to proceed with  some care with the D5-branes, since they can, because of  the
$USp(2Q_{5})$ Chan--Paton factor,  effectively be considered as a pair
of Type-IIB D5-branes. 
Keeping this in mind, we compute the energy of the bound state formed by 
 $Q_{1}$ D1-branes, $Q_{5}$ D5-branes and $N$ units of Kaluza--Klein momentum in
Type-I superstring theory as
\eqn\energ{
M=
{Q_{1}R\over \lambda \alpha^{'}}+{2Q_{5}RV\over \lambda (\alpha^{'})^{3}} 
+{N\over R}
,}
where $R$ is the radius of $S^1_{R}$ and $(2\pi)^4 V$ the volume of the four-torus 
${\bf T}^4$. 
We will see in the following section how this indeed agrees with the ADM mass of the
black hole.

\subsec{Connecting with the black-hole region}

Once we have studied the Type-I black hole from the D-brane side, let us go to 
the strong coupling limit $\lambda Q_{1},\lambda Q_{5},\lambda^{2}N>1$ in which
the D-brane bound state describes a semiclassical black hole with non-trivial  
background values for the metric, dilaton and the Ramond--Ramond two-form.
The solution is formally identical to the one of the five-dimensional Type-IIB 
black hole \refs\rmalda, and it  
is characterized by three radii determining completely the geometry and the 
long-range fields of the black hole (the supersymmetric analogues of the 
Schwarzschild modulus $2G_{N}M\sim \kappa^{2}M$):  
$$
r_{1}^2=\kappa^{2}{Q_{1}(\alpha^{'})^{3}\over  V\lambda}, \hskip 1cm
r_{5}^2=\kappa^{2}{(2Q_{5})\alpha^{'}\over \lambda}, \hskip 1cm
r_{0}^2=\kappa^{2}{N(\alpha^{'})^{4}\over R^2 V}
.$$
The black-hole entropy is defined as 
\eqn\ae{
S={A_H\over 4G_{5}^{2}}
,}
where $A_H$ is the area of the event horizon and $G_{5}$ is the five-dimensional
Newton's constant, proportional to $\kappa^{2}$, the square of the closed string
coupling. In the 
case at hand it can be expressed in terms of $r_{0}^2$, $r_{1}^2$ and $r_{5}^2$ to 
give
\eqn\ech{
S=2\pi\sqrt{NQ_{1}Q_{5}}
,}
which agrees with the corresponding computation from the D-brane side, as it was
noticed in \refs\rdab.
The ADM mass is obtained as well from the large-distance limit of the
 black-hole metric
and coincides with the mass of the D-brane system \energ
$$
M_{ADM}={RV\over \kappa^{2}(\alpha^{'})^{4}}(r_{1}^{2}+r_{5}^{2}+r_{0}^{2})=
{Q_{1}R \over \lambda\alpha^{'}}+{2Q_{5}R V\over \lambda(\alpha^{'})^{3}}+{N\over R}
.$$

\subsec{Enlarging the class of Type-I non-supersymmetric black holes}

It is worth stressing that the extremal
 black hole of  \refs\rdab\
 can be included into a more general class of extremal 
non-supersymmetric black holes constructed by 
replacing the system of D5- and/or D1-branes by the corresponding anti-D-branes.
Beginning with the Type-I {\it supersymmetric} configuration formed by D5-branes, D1-branes
and Kaluza--Klein momentum in the `correct' direction and 
trading, for example, the D1-branes by anti-D1-branes, the reflection condition
for the Killing spinor on the anti-branes gets a minus sign to read $\Gamma_{0}\Gamma_{9}\epsilon=
-\epsilon$. At the same time, the presence of momentum along $X^9$ 
imposes on $\epsilon$ the equation $\Gamma_{0}\Gamma_{9}\epsilon=\epsilon$.
Obviously these two equations are incompatible and consequently no
target-space supersymmetry survives. On the black-hole side this corresponds to
 flipping the 
sign of the charge $Q_{1}$.

What makes this black hole similar to the one studied by Dabholkar is the
fact that it can be related with a supersymmetric one by just reversing 
the sign of the Kaluza--Klein momentum. Doing this, the reflection condition on
the anti-D1-brane is perfectly compatible with the momentum condition on the
Killing spinor and we are left with four unbroken real supersymmetries. Therefore 
the D-brane computation of the entropy can be made along the lines described above.

By trading D-branes with anti-D-branes (i.e. changing the signs of $Q_{1}$ and $Q_{5}$)
we have four possibilities leading to 
non-supersymmetric black holes for a given sign of the momentum along 
$X^9$. Starting from the {\it supersymmetric} Type-I 
black hole made out of D-branes alone, 
denoted by $(Q_1,Q_5,N)$, we find four {\it extremal 
 non-supersymmetric} black-hole solutions
labelled by
$$
(Q_1,Q_5,-N), \hskip 0.8cm 
(-Q_1,Q_5,N), \hskip 0.8cm (Q_{1},-Q_{5},-N), \hskip 0.8cm
(-Q_{1},-Q_{5},N)
,$$
where the first one corresponds to the example constructed in \refs\rdab. In
each case the solution obtained by changing the sign of $N$ is supersymmetric.
Viewing the black holes as bound states of constituent branes, non-supersymmetric
extremality is achieved by combining branes and anti-branes of different
`dimensionality', while non-extremal black holes come from adding some
brane--antibrane pairs.   

More examples can be generated by applying S-dualities to the Type-I black
holes considered here. The whole situation is highly reminiscent of the series of
 extremal non-supersymmetric 
 black holes constructed in refs. \refs\rto.

\newsec{Large $N$ power counting}

In this section we 
 uncover the reasons behind the surprising agreement between
the entropy and mass of D-brane excitations in the weakly coupled version of the Type-I 
black hole 
with the corresponding quantities computed from the low-energy supergravity solution. 
These two regimes are
governed by $\lambda Q_{1}$, $\lambda Q_{5}$ and $\lambda^{2}N$, playing the
role of effective couplings for the open strings degrees of freedom on the D-brane
background.
In both the D-brane and the strongly coupled black-hole picture we work
in the semiclassical limit \scl, the two regimes corresponding to different 
values of the effective couplings \refs\rscl, 
\refs\rdps. At one side we have the D-brane
region in which the open string theory is weakly coupled
$$
\lambda Q_1 < 1, \hskip 1cm \lambda Q_5 < 1, \hskip 1cm \lambda^2 N < 1
,$$
and open-string perturbation theory in the presence of the bound state of D-branes 
is reliable. At the opposite side in the effective coupling moduli space we have
the black hole, for which 
$$
\lambda Q_1 > 1, \hskip 1cm \lambda Q_5 > 1, \hskip 1cm \lambda^2 N > 1.
$$
The physics is that of closed+open strings moving in the background
of the black-hole
 metric, dilaton and R-R antisymmetric tensor. In the $\lambda\rightarrow
0$ limit only genus zero diagrams in the closed-string sector survive.

To visualize the physics behind the success of the D-brane computation of the  
black-hole entropy for the non-supersymmetric Type-I black hole,
 we will study in some 
detail the structure of the Type-I string perturbation theory. In the weakly 
coupled D-brane regime, any observable can be computed in string perturbation 
theory, the corresponding 
Feynman--Polyakov diagrams being Riemann surfaces with an arbitrary number of 
handles $g$, $C$ cross-caps and $B_{1}+B_{5}+B_{9}$ boundaries attached respectively
to the $Q_1$ D1-branes, $Q_5$ D5-branes and the 32 D9-branes characteristic of the
Type-I superstring theory. In addition to this we will have also 
D1--D5 `mixed boundaries' containing insertions of open strings carrying 
Kaluza--Klein momentum\foot{Actually, the Kaluza--Klein momentum is carried by a 
combination of (1,1), (1,5) and (5,5) degrees of freedom, as can be seen 
from the vanishing condition of the flat directions \flat.} along $X^9$. 
The state of the black hole can be specified 
in the Fock space of the $(0,4)$ conformal field theory at the D1--D5 
intersection, by  
giving a set of occupation numbers, $\{n_{ij} (p)\}$,  for strings of momentum
$p$, carrying Chan--Paton labels $(i,j)$ in the fundamentals of 
$SO(Q_1)$ and $USp(2Q_5)$.
We are interested in $S$-matrix amplitudes of the form
\eqn\smat{ \bra \Psi_{BH}', X' |\,S\,|\Psi_{BH}, X\ket,}
where $X$ stands for the quantum numbers of a light system scattered
off the black hole, for example, a single brane-probe, or a set of fundamental
strings. Decay or absorption processes, or mass corrections to the black hole,
can be treated by considering the situation where  either $X$ or $X'$ or both are
trivial.    

A general perturbative amplitude contributing to \smat\
  with $I_o$ external open strings and 
$I_c$ external closed strings is weighted by  
$
\lambda^{-\chi + I_c +I_o /2}
$, 
 where $\chi = 2-2g+B+C$ is the Euler
character. We see that the minimum power of the string coupling is
$\lambda^{-2}$, which scales in the large $N$ limit like\foot{
Given that $Q_1$ and $Q_5$ have the same scaling with $\lambda$, in what
follows $Q$ will stand for either $Q_1$ or $Q_5$. } ${\cal O}(Q^2) \sim 
{\cal O}(N)$, in agreement with the large $N$ scaling of the mass \energ. Therefore,
all diagrams with external closed strings vanish, except for the one- and
two-point functions in the presence of the black hole.  
A closed-string
tadpole is proportional to  $\lambda^{-1} \sim {\cal O}(Q) \sim {\cal O}
(\sqrt{N})$, 
the expected scaling of the Ramond--Ramond fields created by the black hole.
This is also the large $N$ scaling of a brane probe effective action,
since we may substitute the closed-string vertex operator by a single
boundary attached to the probe, with the same coupling dependence.

On the other hand, the closed string form factor, i.e. the two-point function
of closed strings in the black-hole state, is of  
 ${\cal O}(1)$, in agreement with
the fact that it should measure the product $G_N M$.
Since the gravitational field depends on  $G_N M$, we see that the
semiclassical limit is defined in such a way that the weakness of the coupling
constant is balanced by the strength of the sources \refs\rdas.

   By opening D1 and D5 boundaries into a given
Riemann surface, we get extra powers of $\lambda$ paired with the charges 
$Q_1$, $Q_5$ according to $\lambda Q_1$ and $\lambda Q_5$. Unless the 
amplitude vanishes by supersymmetry, these new 
Riemann surfaces do have non-vanishing contributions 
 of  ${\cal O}(1)$ 
in the semiclassical limit. 
Essentially, the weakness of the coupling constant $\lambda$ in the
semiclassical limit is compensated for by the fact that each new boundary can be
attached to either $Q_1$ or $Q_5$ different D-branes, giving rise to $Q_1$ or
$Q_5$ diagrams that contribute coherently. 

A similar enhancement mechanism is at work when considering interactions
between the constituent momentum strings.       
Each boundary with $n$ open-string insertions carries an extra
factor of $\lambda^{1+{n\over 2}}$, but each vertex operator can be attached to
many different strings, so that we get a combinatorial factor from initial-  
and final-state degeneracy in \smat.
This is exactly analogous to the large $N$ power counting rules in
Witten's treatment of baryons \refs\rwittbar.
 For large values of the Kaluza-Klein momentum
$N$, a typical state will share the momentum equally between constituent
strings, of which there are
 ${\cal O}(Q^2)$ species.  
 Therefore, as
 $N/Q^2$ remains fixed in
the semiclassical limit, so does $\langle n_{ij}\rangle$
 in the  open-string condensate, and we can regard the constituent
momentum strings as  ${\cal O}(N)$ species of quarks, 
building a baryon of mass ${\cal O}(N)$. The simplest interactions would then
correspond to gluon exchange between quarks, that is, closed string
exchange between momentum insertions. They are associated to boundaries
with two insertions of the same vertex operator, with a power
$\lambda \cdot (\sqrt{\lambda})^2 \cdot N$, for each boundary, the last
factor being the `quark degeneracy'. 

More complicated interactions involving  more than two insertions per
boundary can also be considered.   
A detailed analysis is carried out in Appendix A, where
we show that only Riemann surfaces with an even number of $(1,1)$, $(5,5)$ or
$(1,5)$  insertions on each boundary survive in the classical limit.
More importantly, we show that the wave-function degeneracy factors
work out in such a way that all corrections scale as a function of
$\lambda^2 N$ at large $N$, and are therefore of  ${\cal O}(1)$. 

The situation is radically different when we add closed-string loops, 
D9-branes holes or cross-caps. In all these cases, the new diagrams are
suppressed by a power $\lambda^{2g+B_9 +C}$ of the string coupling, which 
 is 
now uncompensated by powers of the charges.
Incidentally, this justifies the arguments
of Section 2.1, where we ignored the presence of D9-branes in computing the number 
of flat directions.

The conclusion is that, in the semiclassical regime, $\lambda\rightarrow 0$,
$Q_1,Q_5,N\rightarrow \infty$ with $\lambda Q_1$, $\lambda Q_5$ and $\lambda^2 N$
fixed, all diagrams containing closed-string loops, D9-branes boundaries and
 cross-caps 
will be suppressed by bare positive powers of the vanishingly small string 
coupling constant $\lambda$. In other words, the relevant contributions come only from Riemann 
surfaces with an arbitrary number of D1-brane, D5-brane boundaries, as well as ``mixed'' 
boundaries containing momentum insertions. These surfaces are {\it topologically}
of the same kind as those contributing to the corresponding amplitude in the
supersymmetric Type-IIB superstring black hole. It is important that, as shown in 
Appendix A, no diagram with 
open-string insertions scales like a negative power of $\lambda$, for otherwise we
would be able to get non-vanishing contributions by adding cross-caps or
D9-boundaries. This would spoil the correspondence between the semiclassical
limits of Type-I and IIB black holes.

At a more detailed level, the large $N$ limit of Type-I diagrams involves
Type-IIB diagrams with some numerical rescaling factors. Roughly speaking,
there is a factor of $\half$ for each open-string loop, coming from
the ${\bf Z}_2$ orientation projection in the original Type-I theory. A
detailed analysis in Appendix B shows that such projection factors can
be entirely absorbed in renormalizations of the coupling constants, as well as
the D-brane charges.
In particular, the auxiliary Type-IIB system involves a Type-IIB black
hole of charges $Q_1', Q_5'$ and $N'$, in a vacuum with open- and 
closed-string couplings, $\lambda', \kappa'$ respectively. These parameters
are related to those of the original Type-I system by the dictionary:
$$  
\lambda^{'}={1\over \sqrt{2}}\lambda, \hskip 1cm \kappa^{'}=\kappa
,$$
\eqn\dicc{
Q_{1}^{'}={1\over \sqrt{2}}Q_{1},\hskip 1cm Q_{5}^{'}=\sqrt{2}Q_{5},\hskip 1cm N^{'}=N
.}
Since amplitudes of the Type-I string theory are equal, in the large $N$ limit,  
to the 
corresponding ones for the Type-IIB theory, non-renormalization theorems at work on
the latter 
guarantee the absence of quantum corrections to mass and entropy as we increase
the coupling and pass from the D-brane region to the black hole-domain in the 
non-supersymmetric Type-I black hole.

Microscopically, i.e. from the D-brane point of view, the Type-I D-brane bound state
$(Q_{1},Q_{5},N)$ is equivalent in the semiclassical limit to a Type-IIB D-brane system
characterized by charges $(Q_{1}^{'},Q_{5}^{'},N^{'})$ with mass and entropy
$$
\eqalign{
M^{'} &=
{Q_{1}^{'}R\over \lambda^{'} \alpha^{'}}+{Q_{5}^{'}RV
\over \lambda^{'} (\alpha^{'})^{3}} 
+{N^{'}\over R} \cr
S^{'} &= 2\pi \sqrt{N^{'}Q_{1}^{'}Q_{5}^{'}}
.}
$$
Substituting the primed values in terms of the Type-I ones, we easily find that the
mass of the Type-IIB D-brane bound state equals the mass of the original 
Type-I system, $M^{'}=M$ with $M$ given by \energ. The same happens with the entropy
of D-brane excitations, since $\sqrt{NQ_{1}Q_{5}}=\sqrt{N^{'}Q_{1}^{'}Q_{5}^{'}}$.

On the black-hole side, the geometry of the auxiliary black hole is governed by
the three lengths
$$
r_{1}^{'2}=\kappa^{'2}{Q_{1}^{'}(\alpha^{'})^{3}\over  V\lambda^{'}}, \hskip 1cm
r_{5}^{'2}=\kappa^{'2}{ Q_{5}^{'}\alpha^{'}\over \lambda^{'}}, \hskip 1cm
r_{0}^{'2}=\kappa^{'2}{N^{'}(\alpha^{'})^{4}\over R^2 V}
.$$
Using \dicc,  it is straightforward to check that $r_{i}^{'2}=r_{i}^{2}$ ($i=0,1,5$)
and therefore the geometry of the Type-IIB black hole is identical to the original
non-supersymmetric Type-I black hole. This guarantees that the area of both
event horizons is the same
 as well, and since $\kappa^{'}=\kappa$ the geometrical entropy
of both black holes will coincide and will be equal to the D-brane computation.
The ADM mass, being only a function of $\kappa^{'}$ and $r_{i}^{'2}$, will also
be the same for both black holes and equal to the energy of the weakly coupled 
D-brane system.

\newsec{Near-horizon physics}  

This section is devoted to a more explicit analysis of the large
$N$ limit, as monitored by a near-horizon brane probe \refs\rdps, 
\refs\rmalprob. 
We have seen that 
 the  large $N$ limit
quenches all diagrams involving Type-I cross-caps, $SO(32)$ Chan--Paton
factors, and closed string handles. So we are left with a subset of
orientable diagrams of Type-I string perturbation theory, consisting
of spheres with holes carrying $USp(2Q_5)\times SO(Q_1)$ Chan--Paton factors,
with some
rescaling factors. We have found in particular, for a vacuum amplitude:
\eqn\rel{ \lim_{Q_i \to \infty} \CA (\lambda, Q_1, Q_5)_{\rm I} = 
\lim_{Q_i' \to \infty} \CA(\lambda', Q_1', Q_5')_{\rm IIB}, }
with $Q_1' = Q_1 /\sqrt{2}, Q_5' = \sqrt{2} Q_5$, and $\lambda' = \lambda /
\sqrt{2}$. Such vacuum amplitudes define the loop expansion of the 
world-volume effective action on the branes, and in particular the effective
action for near-horizon brane probes, whose low-energy limit contains a
Dirac--Born--Infeld term.

To be more precise, let us model the black hole-bound state in terms of 
the six-dimensional gauge theory on the D5-branes world-volume. Denoting
by $F_{ab}$ the $USp(2Q_5)$ Yang--Mills field strength, D1-branes bound
to the D5-branes are represented at low world-volume energies as
instanton configurations:
\eqn\instp{Q_1 = {1\over 64\pi^2} \int_{T_4} \varepsilon^{abcd}
\Tr F_{ab} F_{cd},} 
 and the supersymmetry-breaking  Kaluza-Klein momentum in the $X^9$ direction
is introduced
through an appropriate  Yang--Mills configuration such that    
\eqn\mom{
P_9 = \int_{T_4 \times S^1} T_{09} = {1\over g^2} \int_{T^4 \times S^1} 
\Tr \sum_a F_{a0} F_{a9} = {N\over R}, }
where $g^{2}\sim \lambda \alpha'$ denotes the six-dimensional Yang--Mills
coupling.   
 
Finally, we introduce a parallel 
D5-brane probe at a transverse distance $r_i$ and
 velocity $v_i$,   
 where the index $i$ runs over the coordinates transverse
to the D5-branes world-volume. The complete D5 gauge group is now
$USp(2Q_5 +2)$, spontaneously broken to $USp(2Q_5) \times USp(2)$ by
the expectation value of an adjoint scalar charged under both
factors: $U_i = r_i /\alpha'$. If we view the $USp(2Q_5 +2)$ gauge
theory in six dimensions as a dimensional reduction of the corresponding
$\CN=1$ super-Yang--Mills theory in ten dimensions, the adjoint scalars
are given by the transverse components of the ten-dimensional gauge
field $U_i \sim A_i$. Then the relative velocity is proportional
to an electric field in the ten-dimensional theory ${\dot U}_i \sim F_{0i}$. 
More precisely, regarding the probe as fixed and the black-hole bound state
moving with velocity $v_i = \alpha' {\dot U}_i$, we can write 
\eqn\velb{
F_{0i} = {\dot U}_i \Sigma_2
,}
where $\Sigma_2 \equiv \sigma_2 \otimes {\bf 1}_{Q_5}$ is the intertwiner
of $USp(2Q_5)$, relating the complex-conjugated representations
${\overline {\bf 2Q_5}} = \Sigma_2 {\bf 2Q_5} \Sigma_2$.  
   
  The probe effective action in perturbation theory is given by the
low-energy limit of a set of vacuum string diagrams. At $L$ loops we
consider a sphere with $L$ boundaries on the $USp(2Q_5)$ bound state
and one boundary on the probe. The dynamical information on the 
black-hole state is introduced through $USp(2Q_5)$ Wilson lines
\eqn\wl{ W= \Tr \,P\, {\rm exp} \left(i \oint A_{\mu} dx^{\mu} \right)}
at each of the $L$ black-hole boundaries, with $A_{\mu}$ the ten-dimensional
Yang--Mills configuration satisfying conditions \instp, \mom\ and \velb. 
The probe is taken in its ground state, with a trivial Wilson line 
\eqn\wprobe{ W_{\rm pr} = \Tr {\bf 1}_2 = 2. }   
 In order to take the large $N$ limit smoothly, we define the trace-normalized
Wilson lines ${\widetilde W}$ by 
\eqn\trnorm{ W = 2Q_5 \,{\widetilde W}\;,\;\; W_{\rm pr} = 2 \,{\widetilde W}_{
\rm pr},} 
in terms of which the probe effective action takes the form
\eqn\pac{\Gamma_{\rm eff} = \int d^6 x \,
 \CL_{\rm eff} =2 \sum_{\rm surfaces} \lambda^{-\chi} \, (2Q_5)^{B-1} \, 
\bra {\widetilde W}_{\rm pr} 
{\widetilde W}_1 \cdots {\widetilde W}_{B-1} \ket_{\rm I},}
where we have extracted the factor of $2$ from the probe Wilson line. In this
general expression, $\chi$ is the Euler character of the Riemann surface and
$B$ is the total number of holes. In the limit, only spheres with $B= L+1$
boundaries survive, and the Type-I correlation function of Wilson lines is related
to a Type-IIB correlation function via the general identity
\eqn\facgen{ \bra {\widetilde W}_{\rm pr}
 {\widetilde W}_1 \cdots {\widetilde W}_L \ket_{\rm I} \rightarrow
{1\over 2^L} \bra {\widetilde W}_{\rm pr} 
{\widetilde W}_1 \cdots {\widetilde W}_L \ket_{\rm IIB}} and
we obtain
\eqn\prlim{\lim \,
\Gamma_{\rm eff} = \sqrt{2}\sum_{L\geq 0} (\lambda')^{L-1} (Q_5')^{
L} \bra {\widetilde W}'_{\rm pr}
 {\widetilde W}_1' \cdots {\widetilde W}_L' \ket_{\rm IIB} =\sqrt{2} 
\lim \,\Gamma_{\rm eff}',}
with the mapping ${\widetilde W} = {\widetilde W}'$. So, we obtain a Type-IIB
effective action with the rescaled parameters $\lambda', Q_1' , Q_5'$, up to
a global factor of $\sqrt{2}$. Actually, this factor is just the rescaling of
the D5 charge of the probe.   

In order to properly interpret the operator mapping ${\widetilde W}' = {\widetilde
W}$, we make a low-energy expansion in powers of gauge-invariant operators. In
the constant Abelian approximation where we can neglect covariant derivatives
$DF \sim 0$ and commutators $[F,F] \sim 0$, the effective action takes the form
\refs\rmalprob
\eqn\efflow{
\CL_{\rm eff} = \sum_{L\geq 0} \sum_{I\geq 2} C_{L,I} {(\lambda \alpha')^{
L-1} \over U^{2I-2L-4}} \,\sum_{l=0}^{L} (2Q_5)^{L-l} \sum_{n_1 + \cdots n_l =I} 
\Tr F^{n_1} \cdots \Tr F^{n_l} .} 
We have only kept the leading large $N$ operators and have also neglected
higher-dimension operators suppressed by the string scale $\sqrt{\alpha'} U \ll 1$.
In other words, \efflow\ must be understood as a Wilsonian effective action
with an ultraviolet 
 cutoff of the order of $U$. Infrared divergences should be cutoff by
the non-vanishing background field strengths $\bra \Tr F^n \ket \neq 0$.   
The index $I=\sum_i n_i$
 is the total number of gauge-invariant operator insertions, distributed
through $l$ boundaries. The rest of $L-l$ boundaries pick the identity term in
the weak field expansion of the Wilson lines.

Consistency with the approximations of neglecting covariant derivatives and
commutators requires evaluating the effective action on constant Abelian
field strengths of the form
\eqn\ans{ F_{\mu\nu} = {\widetilde F}_{\mu\nu} {\bf H}_{\mu\nu} ,}
with ${\widetilde F}_{\mu\nu}$ representing `group averaged' field
strengths, and 
 ${\bf H}_{\mu\nu}$ matrices in the Cartan subalgebra of $USp(2Q_5)$, whose
trace in the large $N$ limit verifies 
\eqn\cons{ {1\over 2Q_5} \lim_{Q_5 \to \infty} \Tr ({\bf H})^n = 1, \;\;{\rm for }\;
n>1.} 
Then, in this case,  
the statement ${\widetilde W} = {\widetilde W}'$ means  
${\widetilde F}_{\mu\nu} = {\widetilde F}'_{\mu\nu}$, that is, the group-averaged 
field strengths are mapped identically between the Type-I backgrounds and the
auxiliary Type-IIB background. Now, using 
\eqn\trazs{\Tr' F'^2 = Q_5' {\widetilde F}'^2 =
{1\over \sqrt{2}} \cdot 2Q_5 {\widetilde F}^2 = {1\over \sqrt{2}} \Tr F^2 ,}    
we can calculate the renormalization of the Kaluza--Klein charge:
\eqn\kkl{
{N' \over R} = {1\over (2\pi)^3 \alpha' \lambda'} \int \Tr' 
F_N'^2 = 
{1\over (2\pi)^3 \alpha' \lambda} \int \Tr F_N^2 = {N\over R} ,}
and we obtain the expected result $N=N'$, in agreement with the ADM determination
in the previous section. Similarly, Eqs. \instp\ and \cons\ imply 
$Q_{1}'=Q_{1}/\sqrt{2}$.

It has been conjectured that, at least for supersymmetric black holes,
 \efflow\ matches the weak coupling expansion
of the Dirac--Born--Infeld action for a probe propagating in the
near-horizon geometry of the classical black hole \refs\rdps, \refs\rbecks,
 \refs\rmalprob.
 In particular, this
matching requires powerful non-renormalization theorems that reduce
the sum over insertions in \efflow\ to the terms $I= 2L+2$. This is a
powerful constraint. Taking into
account the fact that each loop is suppressed by at least a factor of
$U^{-2}$, from the mass of the stretched strings between the black hole
and the probe,  and using dimensional analysis plus the constraint 
$I=2L+2$, one finds that each boundary contributes 
only two field-strength insertions representing
instantons or momentum, up to higher-dimension operators suppressed by
powers of $\alpha' U^2 \ll 1$.  

We have shown in the previous section that the closed-string backgrounds,
metric, dilaton and RR fields, are exactly matched with the above 
parameter mapping between the original Type-I theory and the auxiliary
IIB vacuum: 
\eqn\mapp{(\kappa, \lambda, Q_1, Q_5 , N) = (\kappa', \sqrt{2} \lambda', 
\sqrt{2} Q_1' , Q_5' /\sqrt{2}, N').}
 Therefore, the Dirac--Born--Infeld
actions are identical up to the renormalization of the overall D5 probe
tension, which appears explicitly in  \prlim. This proves that the    
non-renormalization theorems implied by the conjecture in \refs\rmalprob,
are true of the large $N$ limit of Type-I extremal non-supersymmetric
black holes, provided they hold  for the supersymmetric IIB ones.

So far we have discussed the weak coupling expansion of the effective
action. The dimensionless expansion parameter is the combination  
\eqn\dimex{ g_{\rm eff} =\half  {\lambda \alpha' \bra \Tr F^2 \ket \over U^2} =
 {\lambda' \alpha' \bra \Tr' F'^2 \ket \over U^2}.} 
From the point of view of the large $N$ limit (or, rather, large $Q_5$ limit)
of the $(5+1)$-dimensional theory on the $D5$-branes, this has the form
$$   
g_{\rm eff} = g^2 Q_5 \cdot ({\rm energy})^2
,$$
namely, it is the dimensionless combination of the large $N$ 't Hooft
coupling, and a typical energy in the Yang--Mills theory, in units
of the ultraviolet cutoff, set by $U$. 
Evaluating $\bra \Tr F^2 \ket$ through   \instp, \mom\ and \velb, we obtain
the respective expansion parameters for  velocity, instanton, and momentum
insertions
\eqn\effex{
\eqalign{ g_5 =& {\alpha' {\dot U}^2 \over (2\pi)^2 U^2} \lambda' Q_5' = 
 {\alpha' {\dot U}^2 \over (2\pi)^2 U^2} \lambda Q_5, \cr     
g_1 =& {(2\pi)^3 \alpha' \over V U^2} \lambda' Q_1' = 
 {(2\pi)^3 \alpha' \over V U^2}\cdot\half\cdot \lambda Q_1, \cr
g_p =& {\alpha'^2 \over R^2 V U^2} \lambda'^2 N' =    
{\alpha'^2 \over R^2 V U^2}\cdot \half \cdot \lambda^2 N, }}  
and the series expansions considered above make sense for $g_{\rm eff} \ll 1$. 
Thus, we see that the large $N$ limit can be combined with a low-energy
approximation such that the supergravity description makes sense
for $\lambda Q_1 \sim \lambda Q_5 \sim \lambda^2 N =$ fixed but large, 
provided we keep $\alpha' \times ({\rm energy})^2 $ sufficiently small.
Since $\lambda Q_i$ are the effective expansion parameters of D-brane
perturbation theory, this argument would indicate that the supergravity
description of the near-horizon geometry can be extended to the
region $\lambda Q_i \gg 1$.  A broad generalization of this idea was
recently proposed by Maldacena \refs\rmnew\ to relate the strong-coupling, large $N$
dynamics of conformal theories and supergravity in near-horizon geometries
involving anti-de Sitter factors. In particular, for the case at hand,
the near-horizon geometry of the Type-IIB black hole is given by
$AdS_3 \times T^4 \times S^3$,  the radius of the various factors
being determined in terms of $\lambda, Q_1$ and $Q_5$. Our results imply that
this conjecture  can be immediately generalized to the large $N$ limit
of the conformal field theories relevant to the Type-I black holes,
with the above-mentioned dictionary \mapp\ of parameters $
\lambda' , Q_1'$ and $Q_5'$. 

The conformal field theory in question is the $(0,4)$ superconformal
model that appears as an infrared fixed point of 
the $(1+1)$-dimensional theory of
$(1,1)$, $(1,5)$ and $(1,9)$ strings described in Section 2 (see 
\refs\rcvj\  
for a recent discussion in the present context).        
In this case, at the level of the conformal field theory,
 the large $N$ limit is not of  't Hooft type. Rather, it is the limit
of large central charge $c\rightarrow  {\rm const.} \,\times N^2$. 
The difference between this CFT and the  $(4,4)$ conformal
field theory in the intersection of Type-IIB D5+D1 branes, is simply
the additional sector of  $(1,9)$ strings,
which breaks the supersymmetry of the right movers, and
the ${\bf Z}_2$ orientation projection. The $(1,9)$ sector contributes
only an amount of  ${\cal O}(1)$ to the central charge in the large
$N$ limit, and the orientation projection translates in the dictionary
of parameters that was worked out before. Therefore, the large $N$ limit of
the $(0,4)$ CFT is a certain embedding in a $(4,4)$ CFT with
the usual moduli mapping \mapp.   
                      
This result is exactly analogous to similar
statements involving four-dimensional
non-supersymmetric theories defined as projections
of $\CN =4$ super-Yang--Mills, which exhibit exact conformal invariance
in the large $N$ limit \refs\rvafaetal.     
  
\newsec{Conclusions}

We have studied the Type-I non-supersymmetric black hole  constructed in
ref. \refs\rdab\
and investigated the workings behind the successful D-brane description for this
black hole. Our main result here is that, in spite of dealing
 with a system without
space-time supersymmetry, the supersymmetry-breaking
 contributions to some observables
such as the mass and the entropy are subleading in the large
 $N$ limit at which 
the connection between the D-brane and the black-hole pictures is made.
 We have found
that the Type-I black hole with ${\cal N}=0$ is embedded
 in that limit into a 
Type-IIB black hole with four unbroken supersymmetry charges. Consequently, 
the non-supersymmetric black hole inherits the non-renormalization
 theorems that
guarantee the absence of quantum corrections in the Type-IIB auxiliary theory. 
Actually, this asymptotic restoration of supersymmetry at large
 $N$ in the 
non-supersymmetric Type-I black hole is reminiscent of the idea of
``classical non-renormalization" of the stress-energy
 tensor advocated in \refs\rdab.

This semiclassical equivalence between {\it extremal} Type-I
 and Type-IIB black holes
can be extended to {\it non-extremal} ones since, as we proved 
in Appendix B, the corresponding
perturbation theories are identical in that limit, thus implying
 the coincidence of 
quantum corrections for both theories. This is fully consistent
 with the fact that
Type-I and Type-IIB low-energy
 supergravity black-hole solutions are also identical.

We have extended our study to the dynamics of D5-brane probes
 close to the horizon of
the Type-I black hole and found that, in the semiclassical limit,
 the probe effective action
is identical to the corresponding one for the auxiliary IIB theory.
 Given this large $N$
embedding of the Type-I theory into a Type-IIB one,
 any of the non-renormalization theorems
conjectured for the latter \refs\rmalprob\ must also
 hold for the near-horizon probe
effective action in the Type-I theory in that limit.

Incidentally, the analysis carried out here is also valid for 
the whole class of extremal
non-supersymmetric black holes studied in 
Section 2.3. In each case, in the large $N$ limit the corresponding
 non-supersymmetric
black hole is embedded into an auxiliary Type-IIB theory with
 flipped signs for 
$Q_{1}$ or $Q_{5}$ or both of them. 

\newsec{Acknowledgments}

It is a pleasure to thank I.L. Egusquiza, R. Emparan, A. Feinstein, R. Lazkoz, 
T. Ort\'{\i}n, E. Rabinovici and M.A. Valle-Basagoiti for useful and
interesting discussions. The work of J.L.M. has been partially supported by
the Spanish Science Ministry under Grant AEN96-1668 and by a University of 
the Basque Country Grant UPV-EHU-063.310-EB225/95, and that of M.A.V.-M. by a 
Basque Government Post-doctoral Fellowship.

\appendix{A}{Large $N$ limit of string diagrams with open string insertions}

Here we study the large $N$ scaling of more general open-string interactions,
involving boundaries with more than two momentum insertions.  
As in the text,
 let $n_{ij}$ be the number of $(i,j)$-type open strings. Since the total 
Kaluza--Klein momentum $N$ has to be shared by $Q^2$ species of strings,  
 and $N/Q^2$ remains fixed in
the semiclassical
 limit, so does $\langle n_{ij}\rangle$ in the  open-string condensate. Thus
the number of open strings to which a given vertex operator can be attached will
grow like $Q^2$ at most. Actually, this growth is limited by the Chan--Paton
structure of the interaction.

In order to illustrate this point, let us first consider a Riemann surface where a
single boundary carries four open-string insertions. We have a $\lambda^3$ 
factor, together with the group theory factor
$$ {\rm tr} \Lambda^4=\sum_{ijkl}
\Lambda_{ij}\Lambda_{jk}\Lambda_{kl}\Lambda_{li}.
$$
Naively, the four sums would give rise to $Q^4$ terms, with a contribution 
$\lambda^3 Q^4$ that diverges like $\lambda^{-1}$ in the semiclassical limit.
However, what we really have to consider is the contribution to $S$-matrix
elements, as in \smat, between given initial and final states.  
Consider first the ``diagonal" element
$$
\langle\{n_{ij}(p)\},X'| \,S\, |\{n_{ij}(p)\},X\rangle,   
$$
where the black-hole state is unchanged in the scattering process. Then the two
incoming open-string states taking part in the interaction must be identical
to the two outgoing states. This is achieved by setting $j=l$ or $i=k$. In 
either case we are left with only three sums, and the contribution goes like
$\lambda^3 Q^3$, which is finite in the classical limit.

For off-diagonal elements the situation is different. If $(i,j)$ and $(j,k)$
are the incoming strings, only initial and final states satisfying 
$$
n'_{ij}=n_{ij}-1
, \hskip 1cm n'_{jk}=n_{jk}-1, \hskip 1cm n'_{il}=n_{il}+1, \hskip 1cm
n'_{lk}=n_{lk}+1
$$
can have non-vanishing $S$-matrix elements. And, what is more important, only
{\it one} term in the sum contributes to each of these off-diagonal elements,
which therefore scale like $\lambda^3$ and vanish in the semiclassical limit.
Even if we consider `inclusive' processes where we sum over all possible final
black-hole states, the cross section goes like $\lambda^6 Q^4\rightarrow 0$, 
since we have to sum over {\it probabilities} rather than amplitudes. 

Similarly, for the ``diagonal'' elements of diagrams with a single
boundary with $2k$ insertions, there are $k-1$ constraints on the indices
coming from  the specification of  fixed initial and final states. 
So, we are left with $k+1$ free indices, a power of $\lambda$ from one
boundary, and $(\sqrt{\lambda})^{2k}$ from $2k$ vertex operators, for a total
of $(\lambda Q)^{k+1} \sim (\lambda^2 N)^{k+1 \over 2} \sim {\cal O}(1)$ in 
the large $N$ limit. As is the case with $k=2$, off-diagonal elements can be
seen to vanish in the same limit.

This analysis is easily generalized  to any number of boundaries
with arbitrary insertions. The result is that only diagrams with an even number
of insertions on each boundary give non-zero contributions, and only  to
$S$-matrix elements that are diagonal in the black-hole state. Other
combinations  of insertions and {\it all} off-diagonal elements vanish in the
semiclassical limit.
This is not unexpected, since the fields of a microscopic
string probe vanish for $\lambda  \rightarrow 0$ and are thus unable to affect the
state of the black hole.

\appendix{B}{Type-I versus Type-IIB string perturbation theory}

In this appendix we specify in more precise terms the large $N$ 
 embedding of the Type-I 
 string perturbation series in the black-hole sector, into a similar
Type-IIB system.   
It was argued in Section 3 that
 all diagrams containing closed-string handles, cross-caps or
D9-brane boundaries are suppressed. This implies that both perturbation theories 
(I and IIB) in that limit contain the same type of diagrams from a geometrical point
of view. Since Type-I and Type-IIB theories are close relatives, we will
try to relate the two  theories at a more quantitative level.

Type-I superstring can be obtained from the Type-IIB theory by introducing 
32 D9-branes (a number determined by anomaly cancellation) and projecting out by
the world-sheet parity $\Omega$. This last projection is necessary for the
consistency of the Type-I theory (otherwise we are left with uncancelled dilaton
tadpoles) and results in the introduction of non-orientable Riemann surfaces.
It therefore seems  that, roughly speaking, the main difference between Type-I 
and Type-IIB perturbation theories is the presence in the former of Riemann 
surfaces containing D9-boundaries and cross-caps. If this were just so, both
perturbative expansions would be identical in the
semiclassical limit in which these contributions are suppressed. An 
amplitude with $I_{o}$ symmetric ($\Omega=+1$) open and $I_{c}$ 
closed external string states has a loop expansion of the form
\eqn\peri{
{\cal S}(\kappa,\lambda,I_{o},I_{c})_{\rm I}=
\kappa^{I_{c}}(\sqrt{\lambda})^{I_{o}}\sum_{B=1}^{\infty}
\lambda^{B-2}{\cal A}(B,I_{o},I_{c})_{\rm I}
,}
where $\kappa$ is the closed-string coupling constant, proportional to
the string coupling constant $\lambda$ and the number of loops is $L=B-1$.
For the Type-IIB we have a similar expression
\eqn\periib{
{\cal S}(\kappa^{'},\lambda^{'},I_{o},I_{c})_{\rm IIB}=
(\kappa^{'})^{I_{c}}(\sqrt{\lambda^{'}})^{I_{o}}\sum_{B=1}^{\infty}
(\lambda')^{B-2}{\cal A}(B,I_{o},I_{c})_{\rm IIB}
.}

Although Type-I and Type-IIB perturbation theories contain diagrams of the same kind, 
Type-I diagrams only include string
modes that are even under world-sheet parity inversion, i.e. $\Omega=1$, whereas 
in Type-IIB diagrams those states with $\Omega=-1$ also run in open string loops.
Thus, in order to obtain the Type-I diagrams from its Type-IIB versions we must
impose that only symmetric states under $\Omega$ run in internal channels.
Actually, states of the Type-IIB theory can be classified into symmetric and antisymmetric
with respect to the world-sheet parity inversion $\Omega$; for a state 
$|\alpha,i\bar{j}\rangle$ the symmetric 
and antisymmetric combinations are defined by
$$
|\alpha,ij\rangle_{S(A)}={1\over\sqrt{2}}(1\pm\Omega)|\alpha,i\bar{j}\rangle
.$$
In the closed-string sector, the projection is made  in a similar fashion. States
of the Type-I theory are just the symmetric combinations ($\Omega=1$) of Type-IIB.
Moreover, in an
operator formalism language, we can define the vertex
 $|{\cal V}_{3}\rangle$ corresponding
to a disk amplitude of three open strings. Doing so, we easily find that ($B=1$, 
$I_{o}=3$, $I_{c}=0$)
$$
\eqalign{{\cal A}(1,3,0)\equiv &
\,\langle {\cal V}_3||\alpha_{1},ij\rangle_{S}|\alpha_{2},jk\rangle_{S}|\alpha_{3},
ki\rangle_{S} = 
\langle {\cal V}_3||\alpha_{1},ij\rangle_{S}|\alpha_{2},jk\rangle_{A}|\alpha_{3},
ki\rangle_{A} \cr = & 
\,\langle {\cal V}_3||\alpha_{1},ij\rangle_{A}|\alpha_{2},jk\rangle_{A}|\alpha_{3},
ki\rangle_{S}  =
\langle {\cal V}_3||\alpha_{1},ij\rangle_{A}|\alpha_{2},jk\rangle_{S}|\alpha_{3},
ki\rangle_{A}
,}
$$
whereas any other amplitude containing one or three antisymmetric states
will vanish. Going now to the Type-I theory, external states 
are given by the symmetric combinations of the Type-IIB superstring, and in the sector 
with no D9 holes we have
$$
{\cal A}(1,3,0)_{\rm I}={\cal A}(1,3,0)_{\rm IIB}
.$$
Thus, three-point amplitudes on the disk in Type-I theory are identical to the
corresponding ones in Type-IIB theory. Actually, it is straightforward to 
generalize this result to any amplitude on the disk. By factorization and 
conservation of $\Omega$ on each vertex we find
that
$$
{\cal A}(1,K,0)_{\rm I} =\langle V_{1}|
V_{2}\Delta_{S}\ldots \Delta_{S}V_{K-1}|V_{K}\rangle=
{\cal A}(1,K,0)_{\rm IIB}
,$$
where $\Delta$ is the open-string propagator. Here it is important to stress that 
the vertex operators do not contain any power of the string coupling constant, which 
has been factored out in \peri. The conclusion is that tree amplitudes in the Type-I 
theory are exactly equal to their Type-IIB counterparts.

Let us now introduce loop diagrams. We consider the amplitude of $K$ external
symmetric states in Type-IIB theory. Applying the conservation of $\Omega$
in each interaction, we can write ($B=2$, $I_{o}=K$, $I_{c}=0$)
\eqn\oneloop{ {\cal A}(2,K,0)_{\rm IIB}=
{\rm Tr}_{S}[\Delta V_{1} \Delta\ldots \Delta V_{K}] +
{\rm Tr}_{A}[\Delta V_{1} \Delta\ldots \Delta V_{K}]
,}
where the traces are taken respectively over symmetric and antisymmetric states
running in the loop. But, as we have seen above, the $(SSS)$ 
and $(ASS)$ tree-level 
couplings are equal, so both terms in the right-hand side of \oneloop\ are equal.
Since the first one (the trace over symmetric states) corresponds to the amplitude
for the Type-I theory, we conclude that
$$
{\cal A}(2,K,0)_{\rm I}=
\half {\cal A}(2,K,0)_{\rm IIB}
.$$

The introduction of a closed-string insertion does
 not change in any way the previous results.
Actually, the higher-loop case can be worked out along the lines depicted above. 
Factorizing the amplitude into lower-loop contributions and iterating, we find that
\eqn\gc{
{\cal A}(B,I_{o},I_{c})_{\rm I}=
\left(\half\right)^{B-1} {\cal A}(B,I_{o},I_{c})_{\rm IIB}
,}
with $L=B-1$ the number of open-string loops. The numerical factor appears because
we have to insert a projector onto symmetric states for each open-string loop.

So far we have been working with the individual terms of the perturbative expansions, not
with the resummed series. In order to compare \peri\ with \periib\ we have to know which is
the relation between the Type-IIB couplings $(\lambda^{'},\kappa^{'})$ and
 the original 
ones $(\lambda,\kappa)$. This will be determined by imposing the correct factorization of the
amplitudes. Looking at the annulus amplitude with no open- or
closed-string insertion ($B=2$, $I_{o}=I_{c}=0$) we find from our previous results
\eqn\fac{
{\cal A}(2,0,0)_{\rm I} =\half {\cal A}(2,0,0)_{\rm IIB}
.}
Taking the modulus of the annulus to infinity we find that it factorizes into 
two disk tadpoles joined by a closed-string propagator, namely
$$
{\cal A}(2,0,0)_{\rm I}\longrightarrow 
\left({\kappa\over \lambda}\right)^{2} {\cal A}(1,0,1)_{\rm I}\; \Delta_{C} \;
{\cal A}(1,0,1)_{\rm I}
,$$
and an
 analogous expression for the right-hand side of \fac\ with primed coupling constants.
>From \gc\ we know that ${\cal A}(1,0,1)_{\rm I}=
{\cal A}(1,0,1)_{\rm IIB}$, so using \fac\ we finally have (cf. \refs\rcc)
$$
{\kappa \over \lambda}={1\over \sqrt{2}} {\kappa^{'}\over \lambda^{'}}
.$$
This gives us the Type-IIB couplings in terms of their Type-I counterparts. However,
 this
expression does not determine for us the individual expressions for $\kappa^{'}$ and
$\lambda^{'}$, only their quotient. Here, among all the possibilities, we will take
the most natural one, in which the closed-string
 coupling is the same for both theories
and only the open-string coupling constant transforms
\eqn\transcoup{
\kappa^{'}=\kappa, \hskip 1cm \lambda^{'}={1\over \sqrt{2}}\lambda
.}
Finally we have the following relation between each term of the Type-I and 
Type-IIB
perturbative expansions \peri\ and \periib
\eqn\app{
\kappa^{I_{c}}(\sqrt{\lambda})^{I_{o}+2B-4}{\cal A}(B,I_{o},I_{c})_{\rm I}=
\left({1\over \sqrt{2}}\right)^{B-\half{I_{o}}}
(\kappa^{'})^{I_{c}}(\sqrt{\lambda^{'}})^{I_{o}+2B-4}{\cal A}(B,I_{o},I_{c})_{\rm IIB}
,}
which gives us the correct expression of the large $N$ Type-I amplitudes in terms of
 the Type-IIB ones. It is easy to see that the numerical factor multiplying the
amplitude in \app\ is consistent with the factorization of the amplitudes.

The remaining factors of $1/\sqrt{2}$ in  \app\ can be interpreted as follows.  
For any boundary without open-string insertions, the numerical factor combines
with the Chan--Paton degeneracy into 
$$
{1\over \sqrt{2}} \, {\rm dim} \,(R_{CP}) ,  
$$
where $R_{CP}$ is the representation of the Chan--Paton group of the string
end-points. In our case, D1-branes carry the vector of $SO(Q_1)$, whereas D5-branes
carry 
the fundamental of $USp(2Q_5)$. In Type-IIB, the Chan--Paton group with the rescaled
dimensions  is $U(Q_1')\times U(Q_5')$ with the dictionary of charges in
\dicc. Thus, at least for diagrams without open-string insertions, Type-I amplitudes
in the large $N$ limit
are equal to those in the auxiliary Type-IIB black hole defined by \dicc.

On the other hand, boundaries with open-string insertions do not carry Chan--Paton
degeneracy, but they get combinatorial factors in $S$-matrix elements due to
external-state degeneracy, as explained in  Appendix A. The contribution of
a Riemann surface with a single boundary and $2k$ insertions 
in the Type-I theory will have a factor of $Q^{k+1}$ because of  the degeneracy in 
the external states. According to Eq. \app, the whole amplitude can be written
in terms of the auxiliary Type-IIB theory as
\eqn\prep{
Q^{k+1}
(\sqrt{\lambda})^{2k-2}{\cal A}(1,2k,0)_{\rm I}=Q^{k+1}(\sqrt{2})^{k-1}
(\sqrt{\lambda'})^{2k-2}{\cal A}(1,2k,0)_{\rm IIB}\;.
}
In order to fully relate the amplitudes in Type-I and Type-IIB, we should write $Q$
on the right-hand side of \prep\ in terms of the Type-IIB charges $Q'$, given by
$Q_1'$ or $Q_5'$. 
Using \dicc, we have  $Q=\sqrt{2}Q'$, with $Q$ either $Q_1$ or $2Q_5$, the dimensions  
of the vector representations of $SO(Q_1)$ and $USp(2Q_5)$, respectively.
Putting all together we finally get 
\eqn\final{
Q^{k+1}
(\sqrt{\lambda})^{2k-2}{\cal A}(1,2k,0)_{\rm I}=2^k (Q')^{k+1}
(\sqrt{\lambda'})^{2k-2}{\cal A}(1,2k,0)_{\rm IIB} 
.}
The factor $2^k$ on the Type-IIB side has a simple explanation. On the Type-I
side we have only symmetric $(\Omega=+1)$ open strings as external states; on the
other hand in the Type-IIB auxiliary theory, we have to include both symmetric 
and antisymmetric ($\Omega=-1$)
states when summing over external states 
degeneracy. This
gives us a multiplicity factor of $2$ for each incoming string up to the
total factor $2^k$ appearing in \final. It is easy to see that the above argument
generalizes to any number of boundaries carrying insertions. Therefore we have arrived 
at the conclusion that Type-I and Type-IIB amplitudes with open-string insertions are 
equal in the large $N$ limit.

\listrefs

%\vfill\eject
\bye